\documentclass{article}
\usepackage{spconf,amsmath,amssymb,graphicx}
\usepackage{wrapfig}
\usepackage{balance}
\usepackage{tabularx}
\usepackage{booktabs}
\usepackage{microtype}
\usepackage{caption}
\usepackage{dblfloatfix}
\usepackage[separate-uncertainty=true, multi-part-units=single]{siunitx}
\usepackage{bm}
\usepackage{enumitem}
\usepackage{setspace}
\usepackage{wrapfig}
\usepackage{float}

\usepackage{mwe}
\usepackage{graphbox} %loads graphicx package
\usepackage{filecontents}
\usepackage[hidelinks]{hyperref}
\usepackage{glossaries}
\usepackage{etoolbox}
\newcolumntype{B}{<{\hspace{-1ex}}c}
\usepackage{multirow}
\DeclareSIUnit\px{px}
\DeclareSIUnit\db{dB}
\usepackage[utf8]{inputenc}
\usepackage{csquotes}
\usepackage[style=ieee,backend=biber, url=false,doi=false,isbn=false,eprint=false, maxbibnames=6,minbibnames=1]{biblatex}
\AtEveryBibitem{%
	\clearfield{day}%
	\clearfield{month}%
	\clearfield{endday}%
	\clearfield{endmonth}%
	\clearfield{address}%
	\clearfield{venue}%
	\clearfield{location}%
	\clearfield{publisher}
}

\newcommand{\irow}[1]{% inline row vector
	\begin{matrix}(#1)\end{matrix}%
}
\newcommand{\ignore}[1]{}
\expandafter\def\expandafter\normalsize\expandafter{%
	\normalsize
	\setlength\abovedisplayskip{6pt}
	\setlength\belowdisplayskip{6pt}
	\setlength\abovedisplayshortskip{6pt}
	\setlength\belowdisplayshortskip{6pt}
}

\newacronym{psf}{PSF}{Point Spread Function}
\newacronym{snr}{SNR}{Signal-to-Noise Ratio}
\newacronym{bd}{BD}{Blind Deconvolution}
\newacronym{cnn}{CNN}{Convolutional Neural Network}
\newacronym{map}{MAP}{Maximum a Posteriori}
\newacronym{tv}{TV}{Total Variation}
\newacronym{tvrl}{TV-RL}{Total Variation regularized Richardson-Lucy}
\newacronym{rl}{RL}{Richardson-Lucy}
\newacronym{ssim}{SSIM}{Structural Similarity}
\newacronym[plural=AFs,firstplural=Autofocuses (AFs)]{af}{AF}{Autofocus}
\newacronym{sml}{SML}{Sum of Modified Laplacian}
\newacronym{dnn}{DNN}{deep neural network}
\newacronym{lapv}{LAPV}{Variance of Laplacian}
\newacronym{ewc}{EWC}{Energy of Wavelet Coefficients}
\newacronym{ws}{WS}{Wavelet Sparsity}
\newacronym{gss}{GSS}{Golden Section Search}
\newacronym{sd}{SD}{Standard Deviation} \newacronym{dof}{DOF}{depth-of-field}
\newacronym{hpf}{HPF}{high-pass filter}
\newacronym{na}{NA}{Numerical Aperture}
\newacronym{roi}{ROI}{Region of Interest}
\newacronym{fov}{FOV}{field-of-view}
\newacronym{fwhm}{FWHM}{full width at half maximum}
\newacronym{svm}{SVM}{support vector machines}
\newacronym{mri}{MRI}{magnetic resonance imaging}
\newacronym{gpu}{GPU}{graphics processing unit}
\title{Estimating Nonplanar Flow from 2D Motion-blurred Widefield Microscopy Images via Deep Learning}
\name{Adrian Shajkofci$^{1,2}$, Michael Liebling$^{1,3}$}
\address{$^{1}$Idiap Research Institute, CH-1920 Martigny, Switzerland\\
	$^{2}$\'Ecole Polytechnique F\'ed\'erale de Lausanne, CH-1015 Lausanne, Switzerland\\
	$^{3}$University of California, Santa Barbara, CA 93106, USA}
\begin{document}
%\ninept
%
%\raggedbottom

\maketitle

\begin{abstract}
	Optical flow is a method aimed at predicting the movement velocity of any pixel in the image and is used in medicine and biology to estimate flow of particles in organs or organelles. However, a precise optical flow measurement requires images taken at high speed and low exposure time, which induces phototoxicity due to the increase in illumination power. We are looking here to estimate the three-dimensional movement vector field of moving out-of-plane particles using normal light conditions and a standard microscope camera.
	We present a method to predict, from a single textured wide-field microscopy image, the movement of out-of-plane particles using the local characteristics of the motion blur. We estimated the velocity vector field from the local estimation of the blur model parameters using an deep neural network and achieved a prediction with a regression coefficient of 0.92 between the ground truth simulated vector field and the output of the network. This method could enable microscopists to gain insights about the dynamic properties of samples without the need for high-speed cameras or high-intensity light exposure.
\end{abstract}
\section{Introduction}
\label{sec:intro}
Life is all about movement. From the microscale to the macroscale, organisms undergo growth, nutrients flow or diffuse in their environment. Quantification of the displacement in time of particles, organelles, or organisms can be done using optical flow \cite{corpetti_fluid_2006}. Optical flow is a method aimed at determining the distribution of apparent velocities of any movement in image series. In medicine, and more specifically in cardiac imaging, optical flow proved to be correlated with the flow patterns measured using computational dynamics \cite{brina_intra-aneurysmal_2014}. In photography, \glspl{dnn} recently allowed for the prediction of 3D optical flow in a computationally-efficient way and with a good accuracy \cite{dosovitskiy_flownet_2015, sun_learning_2015, walker_dense_2015, gong_motion_2017,nah_deep_2017, tian_unsupervised_2020}.

Optical flow is usually computed using two image frames at different time points. In microscopy, the physical scales are orders of magnitude smaller than in photography, especially in the axial direction due to the very small depth of field. Consequently, a small movement of the object can cause massive blurring when imaged. For optical flow to be applied successfully, the two reference images must be taken in a short interval of time. High-speed cameras are still uncommon in microscopy stations and fast movement happening during the exposure time causes motion blur. Furthermore, short exposure times require stronger illumination, accelerating phototoxicity or fluorophore depletion \cite{icha_phototoxicity_2017}.

\begin{figure}[t!]
\centering
  \begin{minipage}[b]{1.0\linewidth}
  \centering
\includegraphics[width=0.85\linewidth]{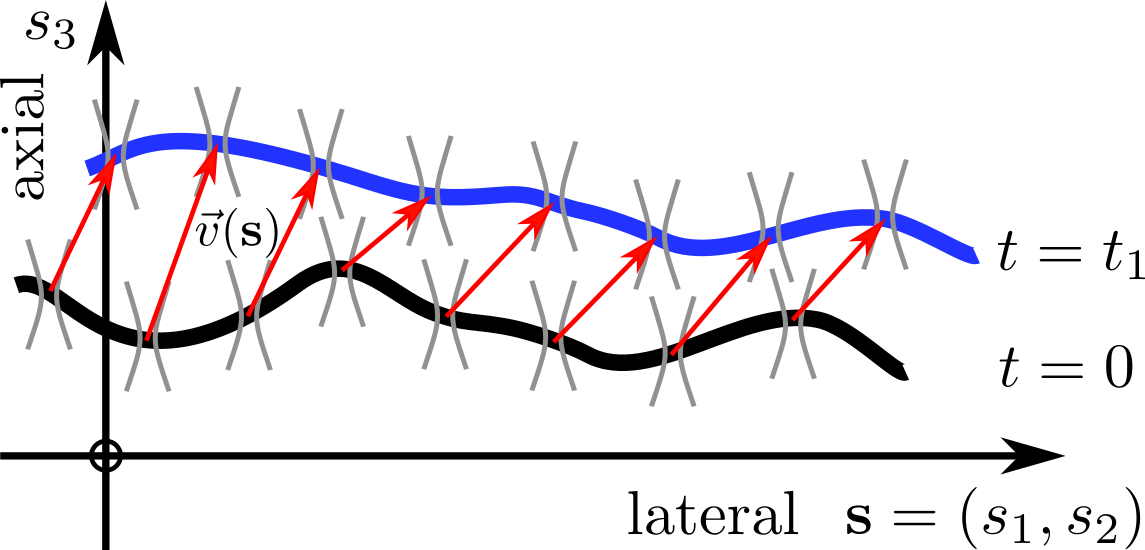}
    \caption*{(a)}
  \end{minipage}
  \begin{minipage}[b]{1.0\linewidth}
  \centering
\includegraphics[width=0.6\linewidth]{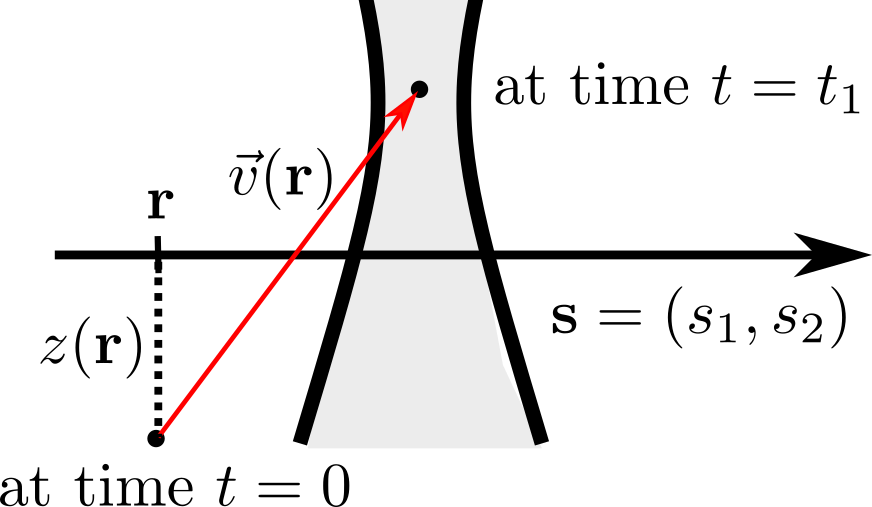}
    \caption*{(b)}
  \end{minipage}
\caption{Image formation model. The object is modeled as a thin moving manifold.  (a) global view of the moving manifold. (b) contribution to image from one point on the manifold at time $t = t_1$; motion is encoded in image blur created by cumulated contributions over exposure interval $[0,t_1]$.}
\label{fig:axis}
\vspace{-1.5em}
\end{figure}

Here we present a method for estimating the movement of out-of-plane particles in a fluid, from a single optical microscopy wide-field image with a long exposure time. We take advantage of the motion blur by estimating the parameters of a spatially-variant \gls{psf} for every point in the image. Since the PSF has been modeled to take into account the displacement in both axial and lateral directions, we are able to extract from the input image a three-dimensional vector field of the motion.

This paper is organized as follows. In Section \ref{sec:methods}, we present the method, comprising the image formation model and the estimation of the displacement vector field. Then, in Section \ref{sec:experiments_flow}, we characterize the performance of the method by firstly simulating arbitrary motion fields in microscopy images, then by simulating a rotational flow of particles in a cylindrical pipe. We then discuss our findings and conclude in Section \ref{sec:experiments_results}.

\section{Methods}
\label{sec:methods}
\subsection{Problem statement}
We consider a single 2D widefield microscopy image $i(\bm{s})$ of a 3D object $o(\bm{s},s_3)$, with $\bm{s}=(s_1,s_2)$ denoting lateral 2D coordinates and $s_3$ the axial coordinate.  We model the object as a flat 2D manifold in a 3D space. We further denote the local movement of the object, measured for points in the image (focus) plane $s_3=z_0$, by a three-dimensional displacement-vector field $\vec{v}(\bm{s}) = \irow{v_1(\bm{s})& v_2(\bm{s})& v_3(\bm{s})}$. We finally define the camera shutter interval $\Delta t$, during which a point in the object initially at position $\bm{s}$ moves to a new position determined by vector $\vec{v}(\bm{s})\Delta t$. Given only the image $i(\bm{s})$  we aim to predict the field of 3D vectors $\vec{v}(\bm{s})$ in the image plane.

\subsection{Image formation model}
\label{sec:formation}
%Similarly to our approach in \cite{shajkofci_semi-blind_2018-1}, 
We assume that the imaged object at time $t=0$ is a thin manifold that can be described as:
\begin{equation}
o(\bm{s},s_3,0) = o_0(\bm{s})\delta\left(s_3-z\left(\bm{s}\right)\right).
\end{equation}
We further assume that the object undergoes a motion of velocity $\vec{v}(\bm{s}) = \irow{v_1(\bm{s})& v_2(\bm{s})& v_3(\bm{s})} = \irow{\bm{u}(\bm{s}) & v_3(\bm{s})}$, with $\bm{u}(\bm{s}) = \irow{v_1(\bm{s}) & v_2(\bm{s})}$.
For the image formation at time $t=t_1$, we start by considering each point on the manifold at time $t=0$, with coordinates $(\bm{r},z(\bm{r}))$ and follow its displacement to position $\left(\bm{r}+\bm{u}(\bm{r})t_1,z(\bm{r})+v_3(\bm{r})t_1\right)$ at time $t=t_1$. This point contributes to the image via a \gls{psf} $h_{\text{3D}}$ centered over it and weighted by the intensity $o_0(\bm{r})$ at $t=0$ (see Fig.~\ref{fig:axis}(b)). Consequently, the intensity at a given position $\bm{s}$ in the image ($s_3=0)$ at time $t_1$ is then given by:  
\begin{equation}
i(\bm{s},t_1) =
\int_{\bm{r}} o_0(\bm{r}) h_{\text{3D}}(\bm{s} - \bm{r} - \bm{u}(\bm{r})t_1, -z(\bm{r})-v_3t_1) \text{d}\bm{r}.
\end{equation}
For a shutter exposure of duration $\Delta t$, we integrate the contributions at each time:
\begin{align}
i(\bm{s}) &= \int_{0}^{\Delta t} i(\bm{s}, t) \text{d}t\\
&= \int_{\bm{r}} o_0(\bm{r}) \underbrace{\int_{0}^{\Delta t}  h_{\text{3D}}(\bm{s} \text{-} \bm{r} \text{-} \bm{u}(\bm{r})t, \text{-}z(\bm{r})\text{-}v_3t) \text{d}t}_{h_{\Delta t}(\bm{s}, \bm{r})} \text{d}\bm{r}\notag\\
&= \int_{\bm{r}} o_0(\bm{r}) h_{\Delta t}(\bm{s}, \bm{r}) \text{d}\bm{r}.
\label{eq:2dconv}
\end{align}
Eq.~(\ref{eq:2dconv}) reveals a spatially variant 2D \gls{psf} $h_{\Delta t}(\bm{s},\bm{r})$, which captures both the local lateral and axial velocities $\bm{u}(\bm{r})$ and $v_3(\bm{r})$ at each location $\bm{r}$ in the image, as well as the local depth of the manifold $z(\bm{r})$. 
This suggests that if the local \gls{psf} could be estimated at every location $\bm{s}$ of an acquired image $i(\bm{s})$, the local 3D velocity field could be estimated, including out-of-plane motion.

\begin{table}[t!]
	\centering
	\caption{Example of predictions of the velocity vector ${\vec{v}}(\bm{s})$. a-c are simulations of beads. e-f are textured images of HeLa cells actin (Alexa Fluor 635) taken with a 10$\times$/0.3 air objective.}
	\begin{minipage}[b]{1.0\linewidth}
	\scalebox{0.85}{
	\centering
	\begin{tabularx}{1.18\linewidth}{llcl}
		 \toprule
			 & ${\vec{v}}(\bm{s})$& $i(\bm{s})$ & $\widetilde{\vec{v}}(\bm{s})$\\\midrule
			 a &
			 	$
			 	\begin{cases}
			 	 {v}_1 = 0.30 \\
			 	 {v}_2 = 0.30 \\
			 	 {v}_3 = 0 \\
			 	 {z}_0 = 0
			 	\end{cases}
			 	$
			  & \includegraphics[align=c, width=0.22\linewidth]{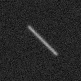} & 
			  			 	$
			  \begin{cases}
			  \widetilde{v}_1 = 0.38 \\
			  \widetilde{v}_2 = 0.31 \\
			  \widetilde{v}_3 = 0.01 \\
			  \widetilde{z}_0 = 0.05
			  \end{cases}
			  $
			  \\
			 b &
				$
				\begin{cases}
				{v}_1 = 0.30 \\
				{v}_2 = 0.30 \\
				{v}_3 = 0.80 \\
				{z}_0 = 0
				\end{cases}
				$
				& \includegraphics[align=c, width=0.22\linewidth]{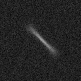} & 
				$
				\begin{cases}
				\widetilde{v}_1 = 0.35 \\
				\widetilde{v}_2 = 0.29 \\
				\widetilde{v}_3 = 0.71 \\
				\widetilde{z}_0 = 0.08
				\end{cases}
				$
				\\
			 c &
				$
				\begin{cases}
				{v}_1 = 0.30 \\
				{v}_2 = 0.30 \\
				{v}_3 = 0.80 \\
				{z}_0 = 0.30
				\end{cases}
				$
				& \includegraphics[align=c, width=0.22\linewidth]{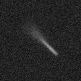} & 
				$
				\begin{cases}
				\widetilde{v}_1 = 0.35 \\
				\widetilde{v}_2 = 0.31 \\
				\widetilde{v}_3 = 0.77 \\
				\widetilde{z}_0 = 0.41
				\end{cases}
				$
				\\			  	
			 d &
				$
				\begin{cases}
				{v}_1 = -0.30 \\
				{v}_2 = 0.60 \\
				{v}_3 = 0.80 \\
				{z}_0 = 0.10
				\end{cases}
				$
				& \includegraphics[align=c, width=0.22\linewidth]{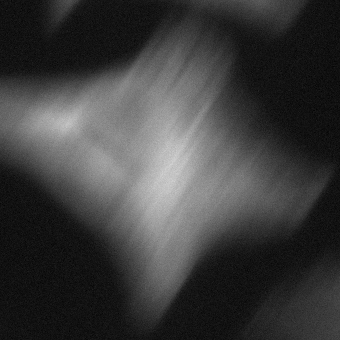} & 
				$
				\begin{cases}
				\widetilde{v}_1 = -0.28 \\
				\widetilde{v}_2 = 0.51 \\
				\widetilde{v}_3 = 0.96 \\
				\widetilde{z}_0 = 0.15
				\end{cases}
				$
				\\			
			 e &
				$
				\begin{cases}
				{v}_1 = 0 \\
				{v}_2 = 0.40 \\
				{v}_3 = 1.20 \\
				{z}_0 = 0
				\end{cases}
				$
				& \includegraphics[align=c, width=0.22\linewidth]{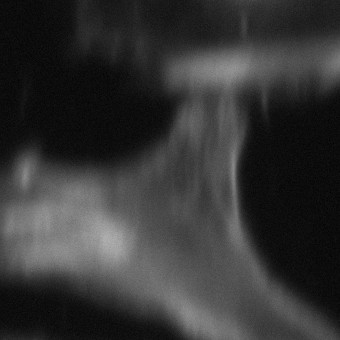} & 
				$
				\begin{cases}
				\widetilde{v}_1 = 0.10 \\
				\widetilde{v}_2 = 0.35 \\
				\widetilde{v}_3 = 0.98 \\
				\widetilde{z}_0 = 0.06
				\end{cases}
				$
				\\														  
			\bottomrule
	\end{tabularx}
}
\end{minipage}
	\label{table1}
\end{table}

\subsection{Estimation of the displacement vector field}
\label{sec:training_flow}
To estimate the displacement vector $\vec{v}(\bm{s})$ from the input image $i(\bm{s})$ we follow a procedure  similar to \cite{shajkofci_spatially-variant_2020}, where we estimated the local \gls{psf} in every location of a still image by training a \gls{dnn} that extracted the \gls{psf} parameters. Here, we again chose for $h_{\text{3D}}$ a Zernike polynomial-based \gls{psf} model \cite{von_zernike_beugungstheorie_1934, shajkofci_spatially-variant_2020}, which we adapted to take into account linear 3D displacement starting at various depths to match Eq.~(\ref{eq:2dconv}).

%We need a displacement estimation function that is invariant to the sample shape or texture but co-variant with the sample's axial position. To that end, we chose an estimator of the local optical properties of the microscope objective that we integrate over time. 

We create a training set of $K = 400'000$ images taken from \cite{zhou_places_2018} that are blurred by spatially-variable \glspl{psf}. To do so, we first define for every $k$-th image, $N$ non-overlapping 2D masks $m_k^n(\bm{s})$, with $n = 0, ... , N-1$. Then, we define, for every mask, a \gls{psf} $h_{\vec{v}^n_k}$ generated using the parameters $\vec{v}_k$ drawn from an uniform distribution ($[-1,1]$ for $(v_1, v_2)$ and $[0,1]$ for $v_3$) and $z_k(\bm{s})$, which is the axial position where the object $o$ is in focus. We get the final $K$ training images by multiplying the masked input image by the \glspl{psf} in the Fourier domain:
\begin{multline}
i_k(\mathbf{s}) = \beta b_p\left(\lambda = \mathcal{F}^{-1}\left[\sum_{n=0}^{N-1}\mathcal{F}(h_{\vec{v}^n_k})\mathcal{F}(m^n * i_k)\right](\mathbf{s}), \bm{s}\right)
\\
+ b_g(\bm{s}),
\end{multline}
with $\beta$ a number between 0 and 1 reflecting the camera quantum efficiency, $b_p(\lambda, \bm{s})$ a random variable following a Poisson distribution, and $b_g(\bm{s})$ a random variable following a zero-mean half-normal distribution. Since there are cases where the \gls{psf} estimation is not possible, e.g. where the sample lacks texture, such as in uniformly black or gray areas, we added a boolean parameter $w_k(\bm{s})$ (whose values can be either $0$ or $1$), which indicates the \enquote{legitimacy} of the sample (i.e is this image textured enough to yield useful information?). We illustrated such input vectors with their corresponding degraded samples in Table~\ref{table1}.

We trained a U-Net \gls{dnn} \cite{ronneberger_u-net_2015} with a ResNet encoder \cite{he_deep_2016} pre-trained on ImageNet \cite{krizhevsky_imagenet_2012}, in order to predict, with the image $i(\bm{s})$ as input, the map of parameters $(\widetilde{\vec{v}}(\bm{s}),\widetilde{z}_k(\bm{s}), \widetilde{w}_k(\bm{s}))$ converted using cylindrical coordinates. We assessed in \cite{shajkofci_spatially-variant_2020} that such a network was robust to unwanted image degradations such as Poisson and Gaussian noise. We trained the network for 50 epochs in PyTorch  with RAdam \cite{liu_variance_2020} optimizing the following loss function:
\begin{multline}
E^{(k)} = \gamma \left({w_k(\bm{s})} - {\widetilde{w}_k(\bm{s})}\right)^2 + \frac {1-w_k(\bm{s})} {U+1} 
\\
\left[\sum_{u=1}^{U}( |v_u(\bm{s})| - |\widetilde{v}_u(\bm{s})| )^2  + (z_k(\bm{s}) - \widetilde{z}_k(\bm{s}))^2\right],
\end{multline}
with $U=3$ components in $\vec{v}$, and $\gamma$ a hyperparameter regulating the importance of the validity parameter $w$, that we set to $1$ in our further experiments.
\section{Experiments}
\label{sec:experiments_flow}
\subsection{Characterization of the displacement vector field estimation with synthetic data}
%\subsection{Characterization of the displacement vector estimation with simulated flow}
We aim at defining the performance of the method using test data generated in the same way as the training data, but with a separate data set of $K_{\text{test}} = 5000$ images cropped at $224 \times 224$ pixel, preliminary acquired using a Leica DM 5500, a 10$\times$/0.3 objective, and fixed fluorescent samples (HeLa cells actin (Alexa Fluor 635) and HeLa cells anti-$\alpha$-catenin (Alexa Fluor 488)). Specifically, we took sharp images of non-moving objects and blurred them with two generated \glspl{psf} modeling different three-dimensional flow rates from a uniform distribution. We then used the \gls{dnn} trained in Section~\ref{sec:training_flow} to predict the flow vector $\widetilde{\vec{v}}(\bm{s})$, the axial position $\widetilde{z}_0(\bm{s})$, and the \enquote{validity} parameter $w_k(\bm{s})$. Since it is a regression problem, our metric was set to be the squared Pearson correlation coefficient $R^2$ averaged over all dimensions.

\begin{figure}
	\centering
	\includegraphics[width=0.8\linewidth]{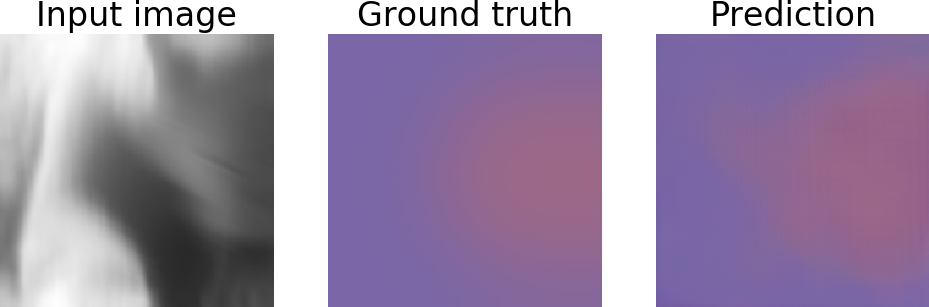}
	\caption{Velocity vector field estimation from a single motion blurred image from the images of fluorescent HeLa cells and two \glspl{psf} drawn from a random distribution. The network output and the ground truth vectors are represented in the RGB spectrum with $\widetilde{v}_1(\bm{s})$ in the red channel, $\widetilde{v}_2(\bm{s})$ in the green channel, and $\widetilde{v}_3(\bm{s})$ in the blue channel. }
	\label{fig:9506122}
\end{figure}
\subsection{Characterization of the displacement vector field estimation with simulated flow}
\begin{figure}
	\centering
	\includegraphics[width=0.85\linewidth]{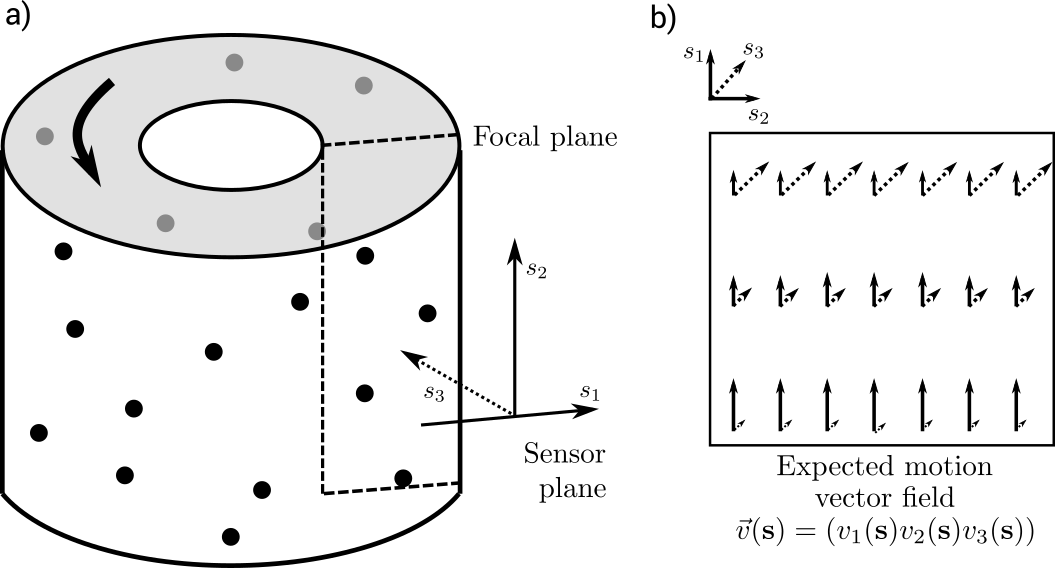}
	\caption{(a) Simulation of a flow in a cylinder. (b) simulation of its expected flow profile as captured by the camera. The flow vector $\vec{v}(\bm{s})$ has a greater lateral component in the bottom of the image, and a larger axial component in the top of the image.}
	\label{fig:simulation_flow}
\end{figure}

We then turned to a more realistic experiment and generated a synthetic testing dataset that simulates the motion of a cylinder where the camera and the focal plane are perpendicular to the motion direction. Due to the small \gls{dof} in microscopy, the effect of the cylinder curvature is negligible. The flow vector map $\vec{v}(\bm{s})$ is then similar to Fig.~\ref{fig:simulation_flow} (b). We neglected as well the effects of the non-slip condition at the walls present in Poiseuille flow.

\begin{figure}
	\centering
	\includegraphics[width=0.8\linewidth]{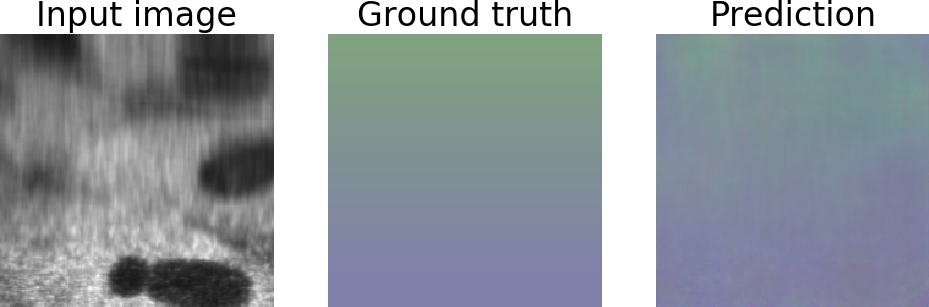}
	\caption{Velocity vector field estimation from a single motion blurred image and a gradient of \glspl{psf} mimicking the conditions of Fig.~\ref{fig:simulation_flow}. The network output and the ground truth vectors are represented in the RGB spectrum with $\widetilde{v}_1(\bm{s})$ in the red channel, $\widetilde{v}_2(\bm{s})$ in the green channel, and $\widetilde{v}_3(\bm{s})$ in the blue channel. }
	\label{fig:9506123}
\end{figure}

%\subsection{Characterization of the displacement vector field estimation with acquired data}

%After having assessed the performance of the method using simulated data, we now turn to the task of the estimation of the displacement vector with actual microscopy acquisitions. 
%To do so, we designed a fluidic system that comprises a pump (Model Pump...), FEP tubing of \SI{200}{\micro\meter} diameter and \SI{20}{\micro\meter} beads at a \SI{1}{\mole / (\micro\meter)^2} concentration. We placed the tubing under the microscope at various angles and ran the pump with a flow rate in the \SI{10}{(\micro\meter)^3 / \second} - \SI{100}{(\micro\meter)^3 / \second} range, corresponding to a velocity in the tube of (I DONT KNOW). The microscope is a custom made bright-field with LED Kohler illumination, 10x apochromat objective (Olympus), and a B\&W camera (Zyla 4.2, Andor).

\section{Discussion and conclusion}
\label{sec:experiments_results}
Our experiments on simulated data confirmed the network's capability to regress a pixel-wise motion vector, not only for point-like sources, but also from a single textured  image blurred with a three-dimensional motion \gls{psf}. Indeed, when it came to the task of estimating two different motion vector in two zones in an image, the network achieved a Pearson regression coefficient of $R^2 = 0.92$ averaged over all pixels of $N=1000$ images of $224 \times 224$ pixel (see Fig.~\ref{fig:9506122}). Similarly, to retrieve the cylindrical flow profile in the second experiment, the regression coefficient was computed at $R^2 = 0.91$ using the same conditions as before (see Fig.~\ref{fig:9506123}).

In all our experiments, the axial component $\widetilde{v_3}$ was predicted with a systematically greater error than the lateral components. That could be explained by the confusion between an object with larger axial velocity that started its motion right in focus, and an object with a smaller axial velocity, but whose motion happens out-of-focus. Both situations yielded similar-looking \gls{psf} since the generation of the \gls{psf} from the parameters $\vec{v}(\bm{s})$ and ${z}_0(\bm{s})$ is not a perfectly bijective transformation (similar-looking \glspl{psf} can be generated from other sets of parameters). However, as illustrated in Table~\ref{table1}, our method for predicting the out-of-plane velocity field was relatively robust to the change in the starting position of the sample $z_0$.

Even though our development is done for thin manifolds, in practice, we expect microscope objectives with sufficiently shallow depths of field and sparse samples to fulfill our method's assumptions. %During the training process, we noticed that the global accuracy was highly sensitive to the training set data size. Indeed, the network showed signs of over-fitting when it was trained with less than $K=100'000$ images. 
We showed that a neural network can be used to predict a motion vector field from a single textured microscopy image degraded with motion blur with only minimal knowledge about the optical setup. This opens the possibility of retrieving flow information without the need for dedicated high-speed camera or strong light exposure.

The source code and trained model is available at the following address:  \url{https://github.com/idiap/flowestimation}.

\balance
\section{Compliance with Ethical Standards}
\vspace{-0.5em}
The post-mortem stained and fixed tissue slices were reused from experiments approved by the EPFL Ethics Committee.
\vspace{-0.5em}
\section{Acknowledgments}
\vspace{-0.5em}
This work was funded by the Swiss National Science Foundation, Grant 200020\_179217.
The authors have no conflict of interest to disclose.

\vspace{-0.5em}
\section{References}
\label{sec:references}
\begin{spacing}{0.9}
	\printbibliography[heading=none]

@article{brina_intra-aneurysmal_2014,
  title = {Intra-{{Aneurysmal Flow Patterns}}: {{Illustrative Comparison}} among {{Digital Subtraction Angiography}}, {{Optical Flow}}, and {{Computational Fluid Dynamics}}},
  shorttitle = {Intra-{{Aneurysmal Flow Patterns}}},
  author = {Brina, O. and Ouared, R. and Bonnefous, O. and van Nijnatten, F. and Bouillot, P. and Bijlenga, P. and Schaller, K. and Lovblad, K.-O. and Grünhagen, T. and Ruijters, D. and Pereira, V. Mendes},
  date = {2014-12-01},
  journaltitle = {American Journal of Neuroradiology},
  volume = {35},
  pages = {2348--2353},
  publisher = {{American Journal of Neuroradiology}},
  issn = {0195-6108, 1936-959X},
  doi = {10.3174/ajnr.A4063},
  url = {http://www.ajnr.org/content/35/12/2348},
  urldate = {2020-08-19},
  eprint = {25082824},
  eprinttype = {pmid},
  langid = {english},
  number = {12}
}

@article{corpetti_fluid_2006,
  title = {Fluid Experimental Flow Estimation Based on an Optical-Flow Scheme},
  author = {Corpetti, T. and Heitz, D. and Arroyo, G. and Mémin, E. and Santa-Cruz, A.},
  date = {2006-01-01},
  journaltitle = {Exp Fluids},
  volume = {40},
  pages = {80--97},
  issn = {1432-1114},
  doi = {10.1007/s00348-005-0048-y},
  url = {https://doi.org/10.1007/s00348-005-0048-y},
  urldate = {2020-08-19},
  langid = {english},
  number = {1}
}

@inproceedings{dosovitskiy_flownet_2015,
  title = {{{FlowNet}}: {{Learning Optical Flow}} with {{Convolutional Networks}}},
  shorttitle = {{{FlowNet}}},
  booktitle = {{{IEEE ICCV}}},
  author = {Dosovitskiy, A. and Fischer, P. and Ilg, E. and Häusser, P. and Hazirbas, C. and Golkov, V. and v d Smagt, P. and Cremers, D. and Brox, T.},
  date = {2015-12},
  pages = {2758--2766},
  issn = {2380-7504},
  doi = {10.1109/ICCV.2015.316},
  eventtitle = {{{IEEE ICCV}}}
}

@inproceedings{gong_motion_2017,
  title = {From {{Motion Blur}} to {{Motion Flow}}: {{A Deep Learning Solution}} for {{Removing Heterogeneous Motion Blur}}},
  shorttitle = {From {{Motion Blur}} to {{Motion Flow}}},
  booktitle = {{{IEEE CVPR}}},
  author = {Gong, Dong and Yang, Jie and Liu, Lingqiao and Zhang, Yanning and Reid, Ian and Shen, Chunhua and Hengel, Anton Van Den and Shi, Qinfeng},
  date = {2017-07},
  pages = {3806--3815},
  publisher = {{IEEE}},
  doi = {10.1109/CVPR.2017.405},
  url = {http://ieeexplore.ieee.org/document/8099888/},
  urldate = {2020-02-11},
  eventtitle = {{{IEEE CVPR}}},
  isbn = {978-1-5386-0457-1},
  langid = {english}
}

@inproceedings{he_deep_2016,
  title = {Deep Residual Learning for Image Recognition},
  booktitle = {{{IEEE CVPR}}},
  author = {He, Kaiming and Zhang, Xiangyu and Ren, Shaoqing and Sun, Jian},
  date = {2016},
  pages = {770--778},
  doi = {10.1109/CVPR.2016.90},
  archivePrefix = {arXiv},
  eprint = {1512.03385},
  eprinttype = {arxiv},
  eventtitle = {{{IEEE CVPR}}}
}

@article{icha_phototoxicity_2017,
  title = {Phototoxicity in Live Fluorescence Microscopy, and How to Avoid It},
  author = {Icha, Jaroslav and Weber, Michael and Waters, Jennifer C. and Norden, Caren},
  date = {2017},
  journaltitle = {BioEssays},
  volume = {39},
  pages = {1700003},
  issn = {1521-1878},
  doi = {10.1002/bies.201700003},
  url = {https://onlinelibrary.wiley.com/doi/abs/10.1002/bies.201700003},
  urldate = {2020-10-25},
  langid = {english},
  number = {8}
}

@incollection{krizhevsky_imagenet_2012,
  title = {{{ImageNet}} Classification with Deep {{Convolutional Neural Networks}}},
  booktitle = {{{NIPS}}},
  author = {Krizhevsky, Alex and Sutskever, Ilya and Hinton, Geoffrey E},
  editor = {Pereira, F. and Burges, C. J. C. and Bottou, L. and Weinberger, K. Q.},
  date = {2012},
  pages = {1097--1105},
  publisher = {{Curran Associates, Inc.}},
  url = {http://papers.nips.cc/paper/4824-imagenet-classification-with-deep-convolutional-neural-networks.pdf},
  urldate = {2017-08-23}
}

@inproceedings{liu_variance_2020,
  ids = {liu\_variance\_2020-2},
  title = {On the {{Variance}} of the {{Adaptive Learning Rate}} and {{Beyond}}},
  booktitle = {{{ICLR}} 2020},
  author = {Liu, Liyuan and Jiang, Haoming and He, Pengcheng and Chen, Weizhu and Liu, Xiaodong and Gao, Jianfeng and Han, Jiawei},
  date = {2020-03-09},
  url = {http://arxiv.org/abs/1908.03265},
  urldate = {2020-04-07},
  archivePrefix = {arXiv},
  eprint = {1908.03265},
  eprinttype = {arxiv},
  langid = {english}
}

@inproceedings{nah_deep_2017,
  title = {Deep Multi-Scale Convolutional Neural Network for Dynamic Scene Deblurring},
  booktitle = {{{IEEE CVPR}}},
  author = {Nah, Seungjun and Kim, Tae Hyun and Lee, Kyoung Mu},
  date = {2017},
  eventtitle = {{{IEEE CVPR}}}
}

@incollection{ronneberger_u-net_2015,
  title = {U-{{Net}}: {{Convolutional Networks}} for {{Biomedical Image Segmentation}}},
  shorttitle = {U-{{Net}}},
  booktitle = {{{MICCAI}} 2015},
  author = {Ronneberger, Olaf and Fischer, Philipp and Brox, Thomas},
  date = {2015},
  volume = {9351},
  pages = {234--241},
  publisher = {{Springer International Publishing}},
  location = {{Cham}},
  doi = {10.1007/978-3-319-24574-4_28},
  url = {http://link.springer.com/10.1007/978-3-319-24574-4_28},
  urldate = {2020-04-07},
  isbn = {978-3-319-24573-7 978-3-319-24574-4}
}

@article{shajkofci_spatially-variant_2020,
  title = {Spatially-{{Variant CNN}}-{{Based Point Spread Function Estimation}} for {{Blind Deconvolution}} and {{Depth Estimation}} in {{Optical Microscopy}}},
  author = {Shajkofci, Adrian and Liebling, Michael},
  date = {2020},
  journaltitle = {IEEE Transactions on Image Processing},
  volume = {29},
  pages = {5848--5861},
  issn = {1941-0042},
  doi = {10.1109/TIP.2020.2986880},
  eventtitle = {{{IEEE Transactions}} on {{Image Processing}}}
}

@inproceedings{sun_learning_2015,
  title = {Learning a {{Convolutional Neural Network}} for Non-Uniform Motion Blur Removal},
  booktitle = {{{IEEE CVPR}} 2015},
  author = {Sun, Jian and Cao, Wenfei and Xu, Zongben and Ponce, Jean},
  date = {2015},
  url = {http://arxiv.org/abs/1503.00593},
  urldate = {2017-09-18},
  archivePrefix = {arXiv},
  eprint = {1503.00593},
  eprinttype = {arxiv},
  eventtitle = {{{IEEE CVPR}} 2015}
}

@article{tian_unsupervised_2020,
  title = {Unsupervised {{Learning}} of {{Optical Flow With CNN}}-{{Based Non}}-{{Local Filtering}}},
  author = {Tian, L. and Tu, Z. and Zhang, D. and Liu, J. and Li, B. and Yuan, J.},
  date = {2020},
  journaltitle = {IEEE Transactions on Image Processing},
  volume = {29},
  pages = {8429--8442},
  issn = {1941-0042},
  doi = {10.1109/TIP.2020.3013168},
  eventtitle = {{{IEEE Transactions}} on {{Image Processing}}}
}

@article{von_zernike_beugungstheorie_1934,
  title = {Beugungstheorie Des Schneidenver-Fahrens Und Seiner Verbesserten Form, Der Phasenkontrastmethode},
  author = {von Zernike, F.},
  date = {1934-05-01},
  journaltitle = {Physica},
  volume = {1},
  pages = {689--704},
  issn = {0031-8914},
  doi = {10.1016/S0031-8914(34)80259-5},
  url = {http://www.sciencedirect.com/science/article/pii/S0031891434802595},
  urldate = {2017-08-23},
  number = {7},
  options = {useprefix=true}
}

@inproceedings{walker_dense_2015,
  title = {Dense {{Optical Flow Prediction}} from a {{Static Image}}},
  booktitle = {{{IEEE ICCV}}},
  author = {Walker, Jacob and Gupta, Abhinav and Hebert, Martial},
  date = {2015-12},
  pages = {2443--2451},
  publisher = {{IEEE}},
  location = {{Santiago, Chile}},
  doi = {10.1109/ICCV.2015.281},
  url = {http://ieeexplore.ieee.org/document/7410638/},
  urldate = {2021-02-05},
  eventtitle = {{{IEEE ICCV}}},
  isbn = {978-1-4673-8391-2},
  langid = {english}
}

@article{zhou_places_2018,
  title = {Places: {{A}} 10 Million Image Database for Scene Recognition},
  shorttitle = {Places},
  author = {Zhou, B. and Lapedriza, A. and Khosla, A. and Oliva, A. and Torralba, A.},
  date = {2018-06},
  journaltitle = {IEEE Trans. Pattern Anal. Mach. Intell},
  volume = {40},
  pages = {1452--1464},
  issn = {0162-8828},
  doi = {10.1109/TPAMI.2017.2723009},
  number = {6}
}
\end{spacing}

\end{document}